\documentclass[twocolumn,reprint,aps,prb,longbibliography]{revtex4-2}
\usepackage[T1]{fontenc}
\usepackage{amsfonts}
\usepackage{comment}
\usepackage{amsmath,amssymb,epsfig,color}
\usepackage{float}
\usepackage{amsmath}
\usepackage{amssymb}
\usepackage{tabularx}
\usepackage{booktabs}
\usepackage{url}
\usepackage{graphicx}
\usepackage{dcolumn}
\usepackage{bbold}
\usepackage{physics}
\usepackage{multirow}
\usepackage{pifont}
\usepackage{siunitx}
\usepackage[normalem]{ulem}

\usepackage{standalone}

\usepackage{hyperref}

\newcommand{\kvec}{\mathbf k}
\newcommand{\nhat}{\hat{\mathbf n}}

\begin{document}

\title{Roughness-robust surface altermagnetism  in $PT$ antiferromagnets}

\author{K. D. Belashchenko}
\affiliation{Department of Physics and Astronomy and Nebraska Center for Materials and Nanoscience, University of Nebraska-Lincoln, Lincoln, Nebraska 68588, USA}
\date{\today}

\begin{abstract}
Surface altermagnetism extends spin splitting beyond bulk altermagnets through symmetry reduction at surfaces and interfaces. An existing classification applies to the local symmetry of atomically flat surface terraces. Here we address the symmetry of macroscopic spin-momentum correlations that survive averaging over compensated rough surfaces. These correlations are governed by the surface antisymmetry Laue class. Because rough-surface altermagnetism is forbidden at any surface of a magnet with antitranslations, the classification reduces to $PT$-symmetric antiferromagnets. By restoring all symmetries leaving the surface normal invariant, roughness can generate compensated surface altermagnets from uncompensated flat terminations or suppress spin splitting. Roughness-robust surface altermagnetism in $PT$-symmetric antiferromagnets offers a route to combine bulk electrical switchability with altermagnetic surface transport.
\end{abstract}

\maketitle

\section{Introduction}

Altermagnets are collinear antiferromagnets whose electronic structure exhibits momentum-dependent spin splitting despite vanishing net magnetization \cite{Smejkal2022a,Mazin2022,Bai2024,Tamang2024,Song_review_2025}. Their combination of spin-polarized transport and negligible stray fields has stimulated considerable interest for spintronic applications \cite{Song_review_2025,jungwirth2025altermagneticspintronics}. In particular, altermagnets can generate spin-polarized currents through the spin-splitter effect \cite{SpinSplitting-Gonzalez,Junwei2021,SpinSplitting-Bai,SpinSplitting-Karube} and often permit electrical readout of the N\'eel vector through the anomalous Hall effect \cite{Smejkal2020}.

Metallic bulk altermagnets remain scarce, especially those with a $d$-wave spin-momentum correlation pattern supporting the spin-splitter effect. One route toward enlarging the set of available materials is provided by surface altermagnetism, where spin splitting emerges through symmetry reduction at a surface of a spin-degenerate bulk antiferromagnet \cite{lange2026emergentaltermagnetismsurfacesantiferromagnets,sasioglu2026dwavesurfacealtermagnetismcentrosymmetric,Hu2026,Leeb2026}.
While surface altermagnets do not naturally generate spin currents flowing perpendicular to the film, they may exhibit altermagnetic in-plane transport properties, including the surface anomalous Hall and nonrelativistic spin-splitter effects.

Symmetry analysis shows that surface altermagnetism can arise at atomically flat, magnetically compensated terraces of various orientations in many antiferromagnets \cite{lange2026emergentaltermagnetismsurfacesantiferromagnets,Hu2026,Leeb2026}.
However, most experimental measurements probe macroscopic responses averaged over rough surfaces containing atomic steps and multiple terminations. This is true both for macroscopic transport devices and for nominally local measurements if the size of the measured area is large compared to the roughness correlation length. For example, the beam diameter in conventional spin-polarized ARPES measurements is typically microns to tens of microns, substantially exceeding typical terrace dimensions in epitaxial films. 

Here we ask which compensated surfaces may exhibit macroscopic altermagnetism after averaging over surface roughness. The nonrelativistic symmetry of the averaged spin-dependent spectral function is obtained by retaining all bulk antisymmetry point group elements compatible with the macroscopic surface normal $\nhat$ \cite{Belashchenko2010,belashchenko2026deterministicelectricalswitchingaltermagnets} and adding effective inversion \cite{Smejkal2022}, which leads to the surface antisymmetry Laue class (SALC). Bulk spin degeneracy at generic wave vectors is enforced either by an antitranslation or by $PT$ symmetry (or both). Antitranslation-protected spin degeneracy is never lifted at a rough surface, leaving only $PT$-symmetric classes available for rough-surface altermagnetism. We derive a complete classification of these rough-surface altermagnets and show that roughness can create surface altermagnetism from uncompensated flat terminations or destroy it by restoring symmetries broken at a flat surface.

\section{Rough-surface altermagnetism and surface antisymmetry Laue classes}
\label{sec:symmetry}

An antisymmetry \cite{Heesch1930} (or black-and-white) point group (APG) or space group (ASG) describes the symmetry of a bipartite crystal, incorporating both its crystallographic symmetry and the assignment of atomic sites to two sublattices. APGs were introduced by Shubnikov \cite{Shubnikov1951} for the classification of bipartite magnetic structures. An element $^{\eta}R\in G$ of an APG combines a real-space symmetry operation $R$ with a symbol $\eta$ denoting whether $R$ maps the two sublattices onto themselves or permutes them; in the modern usage (Litvin notation), $\eta=1$ or $\eta=2$.
APGs provide a natural symmetry description of bipartite magnetic order and its associated response properties and have been used extensively in an equivalent formulation by Turov \cite{Turov1962,Turov1994}.

In strictly collinear magnets, antisymmetry groups are equivalent \cite{Turek2022} to the nontrivial spin groups \cite{Litvin1974}. The nontrivial spin Laue group, obtained by adding effective spatial inversion \cite{Smejkal2022a} to the nontrivial spin point group, governs the presence and angular dependence of nonrelativistic momentum-dependent spin splitting, leading to the definition and classification of three-dimensional altermagnets \cite{Smejkal2022}.

In this paper we consider the symmetry of roughness-robust altermagnetic spin-momentum correlations at macroscopically compensated surfaces of spin-degenerate antiferromagnets. Because roughness breaks microscopic translational invariance, the wave vector $\kvec_\parallel$ is not a conserved quantity. Instead of the electronic band structure, the property of interest is the ensemble-averaged 2D Bloch spectral function $A_\sigma(\kvec_\parallel,\varepsilon)=-\frac{1}{\pi}\langle\mathop{\mathrm{Im}}G_\sigma(\kvec_\parallel,\varepsilon)\rangle$, where $\sigma$ is the spin index, and $G_\sigma(\kvec_\parallel,\varepsilon)$ is the single-particle Green's function, which may be projected on the surface region to account for a specific surface-sensitive probe. The defining property of a rough-surface altermagnet is that its macroscopic symmetry does not enforce spin degeneracy, i.e., $A_\uparrow(\kvec_\parallel,\varepsilon)\neq A_\downarrow(\kvec_\parallel,\varepsilon)$ at a generic $\kvec_\parallel$.

Macroscopic roughness-robust surface responses are controlled by the surface's 2D point group \cite{Belashchenko2010,belashchenko2026deterministicelectricalswitchingaltermagnets}. Indeed, consider a rough surface as an ensemble of microscopic surface configurations with a common macroscopic normal $\nhat$. If the bulk symmetry is lowered only by the presence of a surface normal, surface configurations that can be mapped onto each other by bulk ASG operations should have equal statistical weights in an unbiased ensemble. We therefore assume that the surface ensemble is invariant under all bulk operations preserving the oriented macroscopic surface normal $\nhat$. For example, if the bulk crystal has an antitranslation, any surface termination has a symmetry-equivalent partner with all spins reversed and the crystal shifted by a translation. We will use the term ``rough surface'' in this sense below.

Incorporating sublattice symmetry into the 2D point group gives the surface antisymmetry point group (SAPG), the little group of $\nhat$ in the bulk APG \cite{belashchenko2026deterministicelectricalswitchingaltermagnets}. In two dimensions, the combination of spin rotation and time-reversal symmetries leads to an additional effective symmetry \cite{Smejkal2022}  $A_\sigma(\kvec_\parallel,\varepsilon)=A_\sigma(-\kvec_\parallel,\varepsilon)$, which is a spin-preserving twofold rotation ${}^12_z$ about the surface normal $\nhat$. Adjoining this rotation to the 2D point groups gives the six distinct 2D Laue classes: $2$, $2mm$, $4$, $4mm$, $6$, and $6mm$, and adjoining it to the SAPG gives a surface antisymmetry Laue class (SALC), which forms the basis for our classification.

Spin splitting at a generic $\kvec_\parallel$ point is allowed in non-gray SALCs containing no pure antisymmetry ${}^21$. Requiring magnetic compensation, i.e., that the SALC contains at least one operation with $\eta=2$, results in five altermagnetic SALCs: ${}^2m{}^2m{}^12$ ($d$-wave), ${}^24$ ($d$-wave), ${}^24{}^2m{}^1m$ ($d$-wave), ${}^14{}^2m{}^2m$ ($g$-wave), and ${}^16{}^2m{}^2m$ ($i$-wave). These five SALCs correspond to the five planar-type nontrivial spin Laue groups of Ref. \cite{Smejkal2022}, from which they may be obtained by removing the color-even mirror plane parallel to the surface, which is redundant in the 2D case. The five SALCs also appear in Ref. \cite{le2026catalogaltermagnetismmagneticwallpaperspace} as centrosymmetric 2D altermagnetic point groups. Furthermore, the five altermagnetic SALCs can be formally obtained from the seven centrosymmetric antisymmetry layer point groups \cite{Zeng2024} by restricting spatial operations to their action on the surface plane. This restriction merges two pairs of such layer groups into two SALCs: ${}^22_\parallel/{}^2m$ and ${}^2m{}^2m{}^1m_z$ merge into ${}^2m{}^2m{}^12$, and ${}^1\bar3{}^2m$ and ${}^16/{}^1m{}^2m{}^2m$ into ${}^16{}^2m{}^2m$.

Our consideration is restricted to macroscopic properties that are insensitive to translations. All such properties that are odd under antisymmetry, including the spin asymmetry of $A_\sigma(\kvec_\parallel,\varepsilon)$ at any rough surface, are forbidden in the presence of a bulk antitranslation. If some antitranslations are orthogonal to $\nhat$, antisymmetry-odd responses are forbidden for any atomically flat termination \cite{lange2026emergentaltermagnetismsurfacesantiferromagnets,sasioglu2026dwavesurfacealtermagnetismcentrosymmetric}. If, instead, all antitranslations have a finite component parallel to $\nhat$, antisymmetry-odd responses may be allowed on atomically flat terraces but should vanish after averaging over symmetry-equivalent terraces.

Symmetry-enforced bulk spin degeneracy in the nonrelativistic limit requires the APG $G$ to contain pure antisymmetry ${}^{2}1$, arising from antitranslations in the ASG, anti-inversion ${}^{2}\bar1$, or both \cite{Smejkal2022}. Since antitranslations forbid rough-surface altermagnetism, the remaining candidates have ${}^{2}\bar1\in G$ but ${}^{2}1\notin G$. These are $PT$-symmetric antiferromagnets (PT-AFMs) without antitranslations.

The SAPG is defined as 
$G_{\nhat}=\left\{
{}^\eta R \in G :
R\hat{\mathbf n}=\hat{\mathbf n}
\right\}$. The construction of the SALC $L_{\nhat}$ from $G$ (with ${}^2\bar1\in G$) then leads to a simple rule. Consider an operation ${}^{\eta}R\in G$.
If $R\nhat=\nhat$, then ${}^{\eta}R\in G_{\nhat}$ and hence ${}^{\eta}R\in L_{\nhat}$. If $R\nhat=-\nhat$, then ${}^{\eta}R\notin G_{\nhat}$,
but there is another element ${}^{\bar\eta}\bar R={}^2\bar1\,{}^{\eta}R$ in $G$, where $\bar R=-R$ and $\bar\eta$ is opposite to $\eta$.
Clearly, $\bar R\nhat=\nhat$, and therefore ${}^{\bar\eta}\bar R\in G_{\nhat}$. Multiplying by ${}^12_z$, we obtain an element ${}^12_z\,{}^{\bar\eta}\bar R
={}^{\bar\eta}R'\in L_{\nhat}$, where $R'=m_zR$. 
Thus, the SALC is constructed from bulk operations that preserve or reverse $\nhat$; those that preserve $\nhat$ enter directly with their bulk antisymmetry label $\eta$, and those that reverse $\nhat$ enter multiplied by ${}^1m_z$ with the opposite label $\bar\eta$. For example, an in-plane twofold axis ${}^\eta 2_x\in G$ enters the SALC as ${}^{\bar\eta}m_y$.

\section{Classification of compensated rough surfaces of $PT$-symmetric antiferromagnets}

For the PT-AFMs considered here, a generic surface normal gives $L_{\nhat}={}^{1}2$. A surface whose SALC contains no elements with $\eta=2$ breaks sublattice equivalence \cite{Belashchenko2010}. We will exclude such surfaces from further consideration, because they belong to the ferromagnetic symmetry class, and focus on orientations of $\nhat$ that preserve at least one symmetry operation with $\eta=2$ in the SALC.

Table \ref{tab:ptafm_atlas} enumerates all types of compensated rough surfaces for PT-AFMs, grouped by bulk APG, listing the SALC corresponding to a given $\nhat$. All such surfaces either belong to one of the five altermagnetic SALCs or are gray, i.e., contain pure antisymmetry ${}^21$ and therefore have spin-degenerate spectral functions. Note that the assignment of a given material to a rough-surface SALC depends only on its bulk APG and the surface normal $\nhat$; the translational components of the non-symmorphic elements in the ASG are irrelevant.

\begin{table}[h!]
\caption{\label{tab:ptafm_atlas}
Surface antisymmetry Laue classes (SALCs) for compensated rough surfaces of $PT$-symmetric antiferromagnets with bulk antisymmetry point group $G$ and surface normal $\nhat$ (Cartesian coordinates). Second column: number of corresponding entries in MAGNDATA \cite{MAGNDATA}. The last column lists the altermagnetic spin-splitting type. Gray SALCs enforce nonrelativistic spin degeneracy. For each bulk APG, only one representative is listed for surface orientations with equivalent arrangements of symmetry elements relative to $\nhat$.
}

\begingroup
\setlength{\tabcolsep}{5pt}
\begin{tabular}{lccll}
\toprule
$G$ & Number & $\nhat$ & SALC & Type\\
\midrule

${}^{1}2_y/{}^{2}m$ & 20 & $[h0l]$ &
${}^{2}m{}^{2}m{}^{1}2$ & $d$-wave \\

${}^{2}2_y/{}^{1}m$ & 10 & $[010]$ &
${}^{2}1\,2$ & Gray \\

${}^{2}m_x{}^{2}m_y{}^{2}m_z$ & 3 & [001] &
${}^{2}m{}^{2}m{}^{1}2$ & $d$-wave \\
& & $[hk0]$ &
${}^{2}m{}^{2}m{}^{1}2$ & $d$-wave \\

${}^{1}m_x{}^{1}m_y{}^{2}m_z$ & 63 & $[100]$ &
${}^{2}1\,mm2$ & Gray \\
& & $[hk0]$ &
${}^{2}m{}^{2}m{}^{1}2$ & $d$-wave \\

${}^{1}4_z/{}^{2}m$ & 5 & $[hk0]$ &
${}^{2}m{}^{2}m{}^{1}2$ & $d$-wave \\

${}^{2}4_z/{}^{2}m$ & 4 & [001] &
${}^{2}4$ & $d$-wave \\
& & $[hk0]$ &
${}^{2}m{}^{2}m{}^{1}2$ & $d$-wave \\

${}^{1}4_z/{}^{2}m{}^{2}m_x{}^{2}m_d$ & 0 & [001] &
${}^{1}4{}^{2}m{}^{2}m$ & $g$-wave \\
& & $[100]$ &
${}^{2}m{}^{2}m{}^{1}2$ & $d$-wave \\
& & $[110]$ &
${}^{2}m{}^{2}m{}^{1}2$ & $d$-wave \\
& & $[hk0]$ &
${}^{2}m{}^{2}m{}^{1}2$ & $d$-wave \\
& & $[h0l]$ &
${}^{2}m{}^{2}m{}^{1}2$ & $d$-wave \\
& & $[hhl]$ &
${}^{2}m{}^{2}m{}^{1}2$ & $d$-wave \\

${}^{1}4_z/{}^{2}m{}^{1}m_x{}^{1}m_d$ & 17 & $[100]$ &
${}^{2}1\,mm2$ & Gray \\
& & $[110]$ &
${}^{2}1\,mm2$ & Gray \\
& & $[hk0]$ &
${}^{2}m{}^{2}m{}^{1}2$ & $d$-wave \\

${}^{2}4_z/{}^{2}m{}^{2}m_x{}^{1}m_d$ & 67 & [001] &
${}^{2}4{}^{2}m{}^{1}m$ & $d$-wave \\
& & $[100]$ &
${}^{2}m{}^{2}m{}^{1}2$ & $d$-wave \\
& & $[110]$ &
${}^{2}1\,mm2$ & Gray \\
& & $[hk0]$ &
${}^{2}m{}^{2}m{}^{1}2$ & $d$-wave \\
& & $[0kl]$ &
${}^{2}m{}^{2}m{}^{1}2$ & $d$-wave \\

${}^{2}\bar{3}_z{}^{2}m_y$ & 0 & [001] &
${}^{1}6{}^{2}m{}^{2}m$ & $i$-wave \\
& & $[h0l]$ &
${}^{2}m{}^{2}m{}^{1}2$ & $d$-wave \\

${}^{2}\bar{3}_z{}^{1}m_y$ & 16 & $[010]$ &
${}^{2}1\,2$ & Gray \\

${}^{1}6_z/{}^{2}m$ & 1 & $[hk0]$ &
${}^{2}m{}^{2}m{}^{1}2$ & $d$-wave \\

${}^{2}6_z/{}^{1}m$ & 1 & [001] &
${}^{2}1\,6$ & Gray \\

${}^{1}6_z/{}^{2}m{}^{2}m_y{}^{2}m_x$ & 0 & [001] &
${}^{1}6{}^{2}m{}^{2}m$ & $i$-wave \\
& & $[100]$ &
${}^{2}m{}^{2}m{}^{1}2$ & $d$-wave \\
& & $[010]$ &
${}^{2}m{}^{2}m{}^{1}2$ & $d$-wave \\
& & $[hk0]$ &
${}^{2}m{}^{2}m{}^{1}2$ & $d$-wave \\
& & $[h0l]$ &
${}^{2}m{}^{2}m{}^{1}2$ & $d$-wave \\
& & $[0kl]$ &
${}^{2}m{}^{2}m{}^{1}2$ & $d$-wave \\

${}^{1}6_z/{}^{2}m_z{}^{1}m_y{}^{1}m_x$ & 0 & $[100]$ &
${}^{2}1\,mm2$ & Gray \\
& & $[hk0]$ &
${}^{2}m{}^{2}m{}^{1}2$ & $d$-wave \\

${}^{2}6_z/{}^{1}m_z{}^{1}m_y{}^{2}m_x$ & 0 & [001] &
${}^{2}1\,6mm$ & Gray \\
& & $[010]$ &
${}^{2}1\,mm2$ & Gray \\
& & $[0kl]$ &
${}^{2}m{}^{2}m{}^{1}2$ & $d$-wave \\

${}^{2}m_x{}^{2}\bar{3}_{[111]}$ & 0 & $[001]$ &
${}^{2}m{}^{2}m{}^{1}2$ & $d$-wave \\
& & $[hk0]$ &
${}^{2}m{}^{2}m{}^{1}2$ & $d$-wave \\

${}^{2}m_x{}^{2}\bar{3}_{[111]}{}^{2}m_d$ & 0 & $[001]$ &
${}^{1}4{}^{2}m{}^{2}m$ & $g$-wave \\
& & $[111]$ &
${}^{1}6{}^{2}m{}^{2}m$ & $i$-wave \\
& & $[110]$ &
${}^{2}m{}^{2}m{}^{1}2$ & $d$-wave \\
& & $[hk0]$ &
${}^{2}m{}^{2}m{}^{1}2$ & $d$-wave \\
& & $[hhl]$ &
${}^{2}m{}^{2}m{}^{1}2$ & $d$-wave \\

${}^{2}m_x{}^{2}\bar{3}_{[111]}{}^{1}m_d$ & 7 & $[001]$ &
${}^{2}4{}^{2}m{}^{1}m$ & $d$-wave \\
& & $[110]$ &
${}^{2}1\,mm2$ & Gray \\
& & $[hk0]$ &
${}^{2}m{}^{2}m{}^{1}2$ & $d$-wave \\

\bottomrule
\end{tabular}
\endgroup

\end{table}

The second column in Table \ref{tab:ptafm_atlas} lists the number of entries in the MAGNDATA database \cite{MAGNDATA} corresponding to the bulk APGs, which were assigned using the tabulated collinear spin space groups \cite{Chen_enumeration_SSG}. Several APGs have no or very few known realizations. In particular, none of the currently tabulated magnetic structures belong to the bulk APGs permitting rough-surface $g$-wave or $i$-wave altermagnetism.

Surface symmetry types belonging to gray SALCs in Table \ref{tab:ptafm_atlas} were classified as spin-degenerate surface altermagnets (SD-SAM) in Ref. \cite{lange2026emergentaltermagnetismsurfacesantiferromagnets}, emphasizing that their spin degeneracy does not come from antitranslations. 
The classification of altermagnetic surfaces in Ref. \cite{lange2026emergentaltermagnetismsurfacesantiferromagnets} uses the seven centrosymmetric antisymmetry layer point groups \cite{Zeng2024}, which act on 2D layers embedded in 3D space. As explained in Section \ref{sec:symmetry}, only the action on the 2D plane is relevant for the classification of the nonrelativistic surface spin splitting, which collapses seven centrosymmetric antisymmetry layer point groups into five SALCs.

\section{Discussion}

In contrast to the classifications of Refs. \cite{lange2026emergentaltermagnetismsurfacesantiferromagnets,Hu2026} describing local electronic structure on an atomically flat terrace, the SALCs listed in Table \ref{tab:ptafm_atlas} describe macroscopic spin-dependent spectral functions measured by probes that average over many terraces and terminations. 
Roughness averaging restores all symmetry operations that leave $\nhat$ invariant, regardless of their associated translational components. This symmetry restoration can have two opposite effects. First, some surface orientations of PT-AFMs can be macroscopically compensated and altermagnetic even if the corresponding atomically flat terminations are uncompensated. For example, the tetragonal $P\,{}^24_{2}/{}^2m{}^2n{}^1m$ (e.g., V$_2$WO$_6$) and cubic $F\,^{2}d^{2}\bar3^{1}m$ (e.g., Co$_3$O$_4$) ASGs both have uncompensated atomically flat terminations for $\nhat=[001]$, but their rough $(001)$ surfaces are compensated and have a $d$-wave altermagnetic ${}^24{}^2m{}^1m$ SALC.

Second, the additional symmetries restored by roughness can suppress spin splitting, as in orthorhombic EuZrO$_3$ and CuMnAs (010) (ASG: $P^{1}n^{1}m^{2}a$) or KFeO$_2$ (010) (ASG: $P^{1}b^{1}c^{2}a$). The altermagnetic flat surface of orthorhombic CuMnAs (010) was considered in Ref. \cite{Leeb2026}. In these cases an atomically flat termination loses the sublattice-preserving glide plane perpendicular to the surface, whose glide vector has a component along $\nhat$, but retains the sublattice-swapping one. The resulting flat-surface SALC is $d$-wave ${}^2m{}^2m{}^12$. Once roughness averaging restores the second vertical glide plane with $\eta=1$, the SALC acquires pure antisymmetry and becomes ${}^21mm2$, which is gray.

The APGs of most PT-AFMs allow deterministic electrical switching of the N\'eel vector through bulk staggered current-induced torques \cite{Zelezny2017,Watanabe2018}. Thus, a realization of rough-surface altermagnetism in PT-AFMs may enable the combination of inherent bulk switchability with macroscopic surface altermagnetic responses in the same functional layer.

Unlike bulk altermagnets, altermagnetic surfaces do not naturally generate spin-splitter currents flowing perpendicular to the surface. $d$-wave surface altermagnetism may instead be manifested in an in-plane spin-splitter effect, suggesting lateral transport geometries recently used for Mn$_5$Si$_3$ thin films \cite{mencos2025directdemonstrationtimereversalsymmetrybreakingspin}. For a slab of a $PT$-symmetric antiferromagnet, the termination ensembles representing top and bottom surfaces may, in principle, be related by the $PT$ operation and thereby exhibit opposite spin-momentum correlations. The spin-splitter currents generated at the two surfaces would then be equal in magnitude and opposite in sign. However, such statistical $PT$ symmetry between the two surfaces is usually lifted in heterostructures, which should result in an incomplete cancellation of the spin-splitter currents. In addition, in a lateral edge-injection geometry, the transport layer may be engineered to selectively couple to a particular interface, allowing the spin-splitter effect generated at that surface to be probed.

\section{Conclusion}

Macroscopic spin-momentum correlations at compensated rough surfaces of spin-degenerate antiferromagnets are governed by the surface antisymmetry Laue class (SALC), which depends only on the bulk antisymmetry point group and the surface orientation. Roughness averaging eliminates surface altermagnetism in all magnets with antitranslations and restricts the problem to $PT$-symmetric antiferromagnets. Table \ref{tab:ptafm_atlas} provides a complete classification of macroscopically compensated rough surfaces of $PT$-symmetric antiferromagnets and identifies the corresponding altermagnetic surface classes. This classification describes macroscopic properties emerging after averaging over an unbiased ensemble of surface configurations, in contrast to the local properties of individual atomically flat regions. Surface roughness can create compensated surface altermagnetism from uncompensated flat terminations or cancel out the altermagnetic spin splitting.

\begin{acknowledgments}
I thank the anonymous referee for suggesting the restriction of effective inversion to two dimensions. 
This work was supported by the U.S. Department of Energy (DOE) Established Program to Stimulate Competitive Research (EPSCoR) through Grant No. DE-SC0024284. I used OpenAI GPT 5.6--6 models interactively to explore and refine theoretical arguments and to organize and edit the manuscript. I have critically evaluated and verified all model output used in this work.
\end{acknowledgments}

%\bibliography{refs,biblio,refs2,aps-control}


\end{document}